# GEAMS: A Geographic Energy-Aware Multipath Stream-based Routing Protocol for WMSNs

Samir Medjiah*,**, Toufik Ahmed**, Francine Krief**

*Ecole Nationale Supérieure d'Informatique(ESI),
Ex- Institut National d'Informatique (INI)
BP 68M Oued Smar,
16309 – Alger - Algérie
samir.medjiah@gmail.com

**CNRS-LaBRI,
Université de Bordeaux-1.
351 Cours de la Libération,
33405 – Talence - France
{tad, krief}@labri.fr

**Abstract 1** — Because sensor nodes operate on power limited batteries, sensor functionalities have to be designed carefully. In particular, designing energy-efficient packet forwarding is important to maximize the lifetime of the network and to minimize the power usage at each node. This paper presents a Geographic Energy-Aware Multipath Stream-based (GEAMS) routing protocol for WMSNs. GEAMS routing decisions are made online, at each forwarding node in such a way that there is no need to global topology knowledge and maintenance. GEAMS routing protocol performs load-balancing to minimize energy consumption among nodes using twofold policy: (1) smart greedy forwarding and (2) walking back forwarding. Performances evaluations of GEAMS show that it can maximize the network lifetime and guarantee quality of service for video stream transmission in WMSNs.

## I. Introduction

A wireless sensor network (WSN) [1] consists of light-weight, low power, small size distributed devices called sensor nodes. WSNs have been used in various application areas (military, civil, healthcare, etc.). Examples of applications include forest fire detection, structural health monitoring, target tracking, surveillance… Because of the low node's cost, the deployment of WSN can be ranging from thousand to million nodes, and this can be done randomly or deterministically. A sensor node gathers desired data information, processes it, and transmits it to each other using wireless communication until a base station. The base station (also referred to as the sink node) collects and analyzes the received data from various sensors and draws conclusions about the monitored area. The base station also acts as gateways to other networks.

With the growing-up of miniaturization technology and the availability of low-cost hardware, the sensors nodes embed nowadays various kinds of capturing elements such as simple temperature, microphones, imaging sensors, and video cameras. In this context, the vision of ubiquitous Wireless Multimedia Sensor Networks (WMSNs) [2][3] has become a reality. WMSNs are commonly used for surveillance applications, intrusion detection, target tracking, environmental monitoring, traffic management systems, etc.

These types of applications require addressing additional challenges for energy-efficient multimedia processing, optimal routing and path selection, audio / video rate adaptation to meet the network changing topology, and application specific QoS guarantee (end-to-end delay, loss ratio, data rate…).

Optimal routing in wireless sensor network is a challenging task. Large amounts of research works have been done to enable energy efficiency in WSN. A comprehensive survey of routing protocols in WSN has been presented in [4]. It highlights advantages and performances issues of different routing techniques found in the literature. Despite the similarity between WSNs and MANETs (Mobile Ad-Hoc Networks), routing approaches for MANETs are not suitable to sensor networks. This is due to the different requirements for both networks. First, WSNs contain large number of sensors nodes which augment the communication overhead when MANET protocols are used. Second, the design goal of routing protocols for WSN has to consider energy, power, and storage constraints to maximize the network lifetime and the overall performance. Finally, in WMSN, QoS guarantee in term of low-latency and high reliability data transmission is needed and cannot be met by MANETs protocols.

Routing protocols developed for WMSNs suggest using multipath selection scheme to maximize the throughput of streaming data. Examples of these protocols include: MPMPS (*Multi-Priority Multi-Path Selection*) [5] and TPGF (*Two-Phase Geographical Greedy Forwarding*) [6]. However, such protocols have to build a complete map of the network topology to select the optimum routing / transmission path between the source and destination. They are not adapted in large-scale, high density and frequent mobility situations. Hierarchical routing is the most adopted approach to scale to large network. The creation of clusters with different capabilities can greatly contribute to overall system scalability, lifetime, and energy efficiency. However, the overhead can increase dramatically when network topology changes frequently. In particular, as nodes die and leave the network, update and reconfiguration mechanisms should take place to update the cluster or to select a new cluster head.

This operation overhead will favor the dying of new nodes. Examples of these protocols include LEACH [7], PEGASIS [8]. Geographical routing can achieve scalability in WSNs. GPSR (*Greedy Perimeter Stateless Routing*) [9] was defined to increase network scalability under large number of nodes. The advantage is that the propagation of topology information is required only for a single hop. However, greedy forwarding relays on local-knowledge in which always best node to destination is selected. There are situations in which only a particular path to destination is preferred (for example a path with the minimum transmission delay). In such a case, selecting the same path will lead to premature dying of nodes along this path.

In this paper, we examine the benefit of geographical routing along with multipath local-based route selection and we propose a new routing algorithm namely GEAMS (a Geographic Energy-Aware Multipath Stream-based) routing protocol that leverage both energy constraint and QoS sensitive stream such as audio and video.

The rest of this paper is organized as follow. To make this paper self readable, we expose in section II the routing protocols that influenced the design of GEAMS such GPSR and MPMPS. In section III, we present the functionalities of GEAMS protocol. In section IV, the performance evaluation of GEAMS will be presented. Section V will conclude this paper.

## II. RELATED WORK

Geographical routing sheds light upon the process in which each node is aware of its geographic positing and uses packet's destination address as a geographic position to perform routing and forwarding decision. Since the communication between the source and the destination nodes may require traversal of multiple hops, it is therefore essential to maximize the lifetime of the network and to minimize the power usage of each node in order to assure an optimal routing decision. Two important protocols have been defined that make use of node positing for packet forwarding decision: GPSR and MPMPS. MPMPS is itself based on TPGF. These protocols are briefly described in what follows.

### A. GPSR

The GPSR (*Greedy Perimeter Stateless Routing*) [9] was originally designed for MANETs but rapidly adapted for WSNs. The GPSR algorithm relies on the correspondence between the geographic location of nodes and the connectivity within the network by using the location position of nodes to forward a packet. Given the geographic coordinates of the destination node, the GPSR algorithm forwards a packet to destination using only one single hop location information. It assumes that each node knows its geographic location and geographic information about its direct neighbors.

This protocol uses two different packet forwarding strategies: *Greedy Forwarding* and *Perimeter Forwarding*. When a node receives a packet destined to a certain node, it chooses the closest neighbor out-of itself to that destination and forwards the packet to that node. This step is called the *Greedy Forwarding*. In case that such node cannot be found, (i.e. the node itself is the closest node to the destination out-of its neighbors but the destination cannot be reached by one hop), the *Perimeter Forwarding* will be used. The *Perimeter Forwarding* occurs when there is no neighbor closest to Destination (D) than node (A) itself. Figure 1 illustrates that node A is closer to D than its neighbors x and y. This situation is called "voids" or holes. Voids can occur due to random nodes deployment or the presence of obstacles that obstruct radio signals. To overcome this problem, *Perimeter Forwarding* is used to route packets around voids. Packets will move around the void until arriving to a node closest to the destination than the node which initiated the *Perimeter Forwarding*, after which the *Greedy Forwarding* takes over.

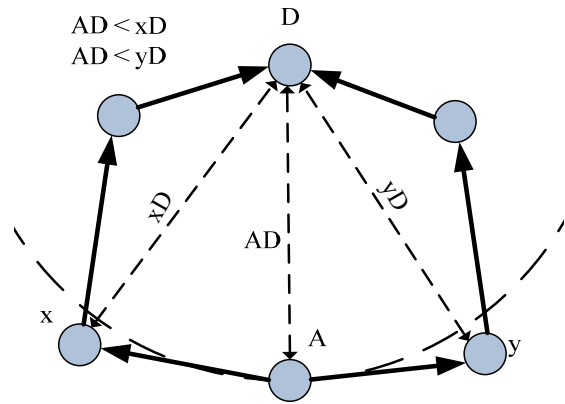

**Figure 1: GPSR Perimeter forwarding to bypass a void or a hole.**

By maintaining only information on the local topology, the GPSR protocol can be suitable for WSNs. However, the greedy forwarding leads to choose only one path from the source to the destination.

### B. TPGF

TPGF (*Two Phase geographical Greedy Forwarding*) [6] routing protocol is the first to introduce multipath concept in wireless multimedia sensor networks (WMSNs) field. This algorithm focuses in exploring and establishing the maximum number of disjoint paths to the destination in terms of minimization of the path length, the end-to-end transmission delay and the energy consumption of the nodes.

The first phase of the algorithm explores the possible paths to the destination. A path to a destination is investigated by labeling neighbors nodes until the base station. During this phase, a step back and mark is used to bypass voids and loops until successfully a sensor node finds a next-hop node which has a routing path to the base station. The second phase is responsible for optimizing the discovered routing paths with the shortest transmission distance (i.e. choosing a path with least number of hops to reach the destination). The TPGF algorithm can be executed repeatedly to look for multiple node disjoint-paths.

### C. MPMPS

The MPMPS (*Multi-Priority Multi-Path Selection*) [5] protocol is an extension of TPGF. MPMPS highlights the fact that not every path found by TPGF can be used for transmitting video because a long routing path with long end-to-end transmission delay may not be suitable for audio/video streaming. Furthermore, because in different applications, audio and video streams play different roles and the importance level may be different, it is better to split the video stream into two streams (video/image and audio). For example, video stream is more important than audio stream in fire detection because the image reflects the event, audio stream is more important in deep ocean monitoring, while image stream during the day time and audio stream during the night time for desert monitoring. Therefore, we can give more priority to the important stream depending on the final application to guarantee the using of the suitable paths.

### D. Discussion

Generally, a WSN is covered by densely deployed sensor nodes. Knowing the full map (network topology) of the deployed nodes in the network to perform routing as done by TPGF and MPMPS is not suitable for many reasons: (1) the exchange of the network map is energy consuming, (2) the exchanged map may not reflect the actual topology of the network, (3) nodes mobility and nodes dying are more frequent in WSN than in other ad hoc networks. These reasons are valid when paths are selected *a priori* by protocols such as TPGF and MPMPS. In these protocols, the selected path is chosen in advance from the source to the destination based on route discovery mechanisms which run before the delivery phase. However, the actual map of the network may change. The GPSR protocol forwards the packet hop by hop based on local available information. This protocol seems to be more promising to scale to large network but does not achieve load balancing in a statistical sense and by making use of multipath routing in WSNs.

In this paper, we propose a new geographical routing protocol namely GEAMS (*Geographic Energy-Aware Multipath Stream-based*) that routes information based on GPSR functionalities (*Greedy Forwarding* and *Perimeter Forwarding*) while maintaining local-knowledge for delivering this information on multipath basis.

### III. GEAMS ROUTING PROTOCOL

The GEAMS routing protocol can be seen as an enhancement of the GPSR protocol to support the transmission of video streams over wireless sensor networks. The main idea is to add a load-balancing feature to GPSR in order to increase the lifetime of the network and to reduce the queue size of the most used nodes. In fact, routing with GPSR will always choose the same path (i.e. using the same node which is closer to destination). This will rapidly cause the dying (dropping) of the most used nodes. In GEAMS routing protocol, data streams will be routed by different nodes, decisions are made at each hop avoiding the algorithm to maintain a global knowledge of the topology.

The design of GEAMS was driven by the following points:
- *Shortest path transmission:* multimedia applications generally have a delay constraint which requires that the multimedia streaming in WSNs should always use the shortest routing path which has the minimum end-to-end transmission delay. Using the same path to a particular destination as done by GPSR will increase the queuing size of the nodes along the transmission path. This will affects considerably the end-to-end transmission delay.
- *Multipath transmission:* Packets of multimedia stream are generally large in size and the transmission requirement can be several times higher than the maximum transmission capacity of sensor nodes. To boost the transmission capacity of the source, it is essential to make use of parallel multipath connection across the available paths.
- *Load balancing*: because of the density of a WSNs, a load balancing feature during the design of a routing protocol has to be considered to avoid frequent node failures and consequently to maximize the network lifetime.

Depending on the data delivery model (Event-Driven, Query-Driven …), the source node splits an image into small packets. Packets are forwarded from one node to another node until reaching the destination node according to a certain policy.

At each hop, a forwarder node decides through which neighbor it will send the packet. Decision policy at each node is based on these four rules: (1) the remaining energy at each neighbor, (2) the number of hops made by the packet before it arrives at this node (3), the actual distance between the node and its neighbors, and (4) the history of the packets forwarded belonging to the same stream.

The GEAMS routing protocol has two modes, the *Smart Greedy Forwarding* and the *Walking Back Forwarding*. The first mode is used when there is always a neighbor closer to the sink node than the forwarder node, while the second one is used to get out of a blocking situation in which the forwarder node can no longer forward the packet towards the sink node. The following section will explain the two routing modes.

### A. Smart Greedy Forwarding:

In GEAMS routing protocol, each sensor node stores some information about its *one*-hop neighbors. Information includes the estimated distance to its neighbors, the distance of the neighbor to the sink, the data-rate of the link, and the remaining energy. This information is updated by the mean of beacon messages, scheduled at fixed intervals. Relaying on this information, a forwarder node will give a score to each neighbor according to an objective function "f(x)". (See Figure 2)

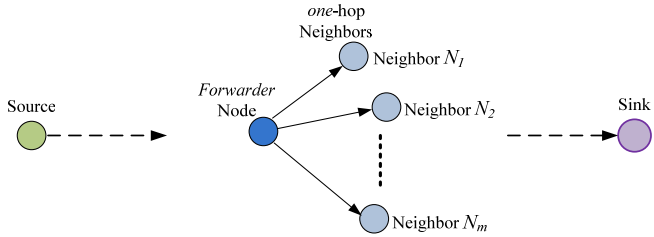

**Figure 2: One-hop neighbors sorted by their score.**

*Packet energy consumption*

When a node (A) sends a packet (pk) of n bits size to a node (B), as illustrated in Figure 3, the energy of node A will decrease by $E_{TX}(n, \overline{AB})$ while the energy of the node $B$ will decrease by $E_{RX}(n)$. Consequently, the cost of this routing decision is $E_{TX}(n, \overline{AB}) + E_{RX}(n)$ considering the energy of the whole network.

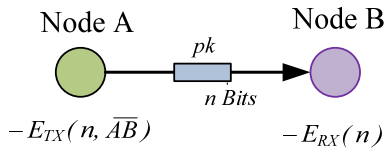

**Figure 3: Packet energy consumption.**

We assume that the transmitted data packets in the network have the same size. We propose an objective function to evaluate a neighbor $N_i$ for packet forwarding. This objective function takes into account the packet energy consumption and also the initial energy of that neighbor. The proposed objective function can simply be:

$$f(N_i) = N_{i_{Energy}} - E_{TX}(N_{i_{Distance}}) - E_{RX}$$

Where: $E_{TX}(D)$ is the estimated energy to transmit a data packet through a distance D, and $E_{RX}$ is the estimated energy to receive the data packet.

These two functions rely on the energy consumption model proposed by *Heinzelman et al.* [7]. According to this model, we have:

$$E_{TX}(k, D) = k \cdot (E_{ELEC} + \varepsilon_{amp} \cdot D^2)$$
$$E_{RX}(k) = k \cdot E_{ELEC}$$

Where:
**k** is the size of the data packet in bits,
**D** is the transmission distance in meters,
$E_{ELEC}$ is the energy consumed by the transceiver electronics,
$\varepsilon_{amp}$ is the energy consumed by the transmitter amplifier.
$E_{ELEC}$ was taken to be 5 $\mu$J/bit and $\varepsilon_{amp}$ 1 $\eta$J/bit.

For each known source node $s_i$ a forwarder node ($N$) maintains a couple $(H_i, j)$. $H_i$ represents the mean hopcount that separates $s_i$ to $N$, and $j$ represent the neighbor whom score is closest to the average score of all closest nodes to the sink. Upon receiving a data packet from the source node $s_i$, the forwarder node will retransmit the packet to a neighbor that is closest to the sink node and in such a way that the number of hops the packet did, will meet the rank of that neighbor. The main idea is to forward a packet with the biggest number of hops through the best neighbor, consequently a packet with the smallest number of hops through the worst neighbor to allow best load balancing in the network. The following algorithm describes the forwarding policy. Figure 5 and Figure 6 present the two scenarios.

---

***Upon_Recieving_a_Packet*** ( *pk* )

*Inputs:*

   **Best_Neighbor**: *a set of the closest neighbors to the sink node sorted in descending order by their score {BN₁, BN₂, ... BNₘ}.*

   **m = |Best_Neighbor|. m** represents the cardinal of the Best_Neighbor set

   **j** :index of the node in the set **Best_Neighbor** whom score is closest to the average score of all closest nodes to the sink. For example, if **Best_Neighbor is {8,5,2,1}** the average score is **4** then **j=2** (starting from index=1)

*Utilities:*

   **Get_Hop_Values** (S$_i$) returns the stored values of empirical hop count from already known source S$_i$ and the j index of the average score of all closest nodes to the sink. These values are (H$_i$, j)

   **Set_Hop_Values** (S$_i$, H$_i$, j) sets the empirical hop count for source S$_i$ to be H$_i$ and j to be the index of the average score of Best_Neighbor set.

   **Forward** (pk, BN$_k$ ) forwards the packet **pk** to the neighbor **k** which has BN$_k$ score

---

```
1:   if (Get_Hop_Values (pk.SourceNode) is Null ) {
2:       Forward (pk, BN₁)        // Default forward to best node
3:       H ← pk.HopCount
4:       Set_ Hop_Values (pk.SourceNode, H, j)
     }
5:   else {      //Get_Hop_Values (pk.SourceNode) is not null
6:       (H,j) ← Get_Hop_Values (pk.SourceNode)
7:       Δh ← H – pk.HopCount
8:       index ← j + Δh
9:       case (index ≤ 0) {
10:          H ← H–index +1
11:          index←1  // index of the best node in neighbor_Set
12:      }
13:      case ( index > m ) {
14:          H ← H–index+m
15:          Index ←m //index of the worst node in neighbor_Set
16:      }
17:      Forward ( pk, BN_index ) // Smart forward
18:      Set_ Hop_Values ( pk.SourceNode, H,j)
19: }
```

**Figure 4: the Smart Greedy Forwarding algorithm.**

Line 1 allows checking if we have already received a packet from an already known source node. If no, the packet will be always forwarded to the best node (line 2), and we have to save the hop count "H" and the average score index "j" in the best neighbor set. These empirical values will be used later to allow load balancing. Figure 5 illustrates the forwarding of the packet to the best node (index=1).

It is clear that the first packet received from an unknown source will be always forwarded to the best neighbor node. According to line 1, if the packet is received from an unknown source, thus the "Get_Hop_Values" will return a null value. The packet will be forwarded to the best node (BN₁) according to line 2.

Line 5 specifies that we have already an empirical estimation of the hop count H and the average index j from a particular source. These values are retrieved in line 6. We calculate in line 7, the deviation Δh of the hop count of the received packet compared to the stored value H. The index of the new forwarder neighbor that allows best load balancing will be adjusted by Δh (line 8). However, two different out of range situations may occur. Line 9 specifies that the received packet has experienced a lot of hops, and thus it needs to be forwarded later to the best node (i.e. node with index=1). In line 13, the received packet has experienced a less hop count than the empirical value H, and thus it has be forwarded to node with higher index (index=m). Line 10 and line 14 compute the new empirical value that will be used later as a new reference. Therefore, the smart forwarding occurs in line 17.

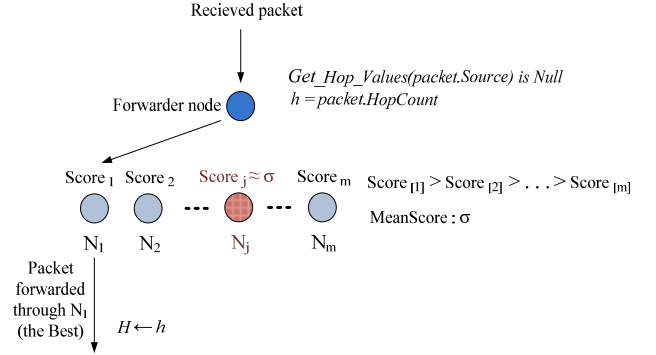

**Figure 5: Forwarding the first packet of a data stream.**

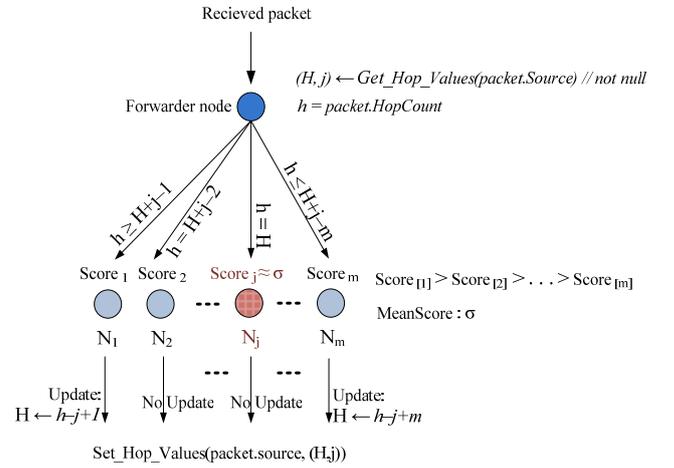

**Figure 6: Forwarding a packet of an already known stream.**

### B. Walking Back Forwarding

Because of node failure, node mobility, and node activity and scheduling policy, disconnections may occur in a WSNs generating what we call "voids". At a certain stage, a forwarder node may face a void where there is no closest neighbor to the sink as illustrated in Figure 7. In this case, the node enters the walking back forwarding mode in order to bypass this void.

In such a case (see Figure 7), the forwarder node will inform all its neighbors that it cannot be considered as a neighbor to forward packets to the sink. This node will also delegate the forwarding responsibility to the less far of its neighbors to bypass the void. This process is recursively repeated steps back until finding a node which can forward successfully the packet.

Figure 8 presents an overview diagram of GEAMS Routing mode switching.

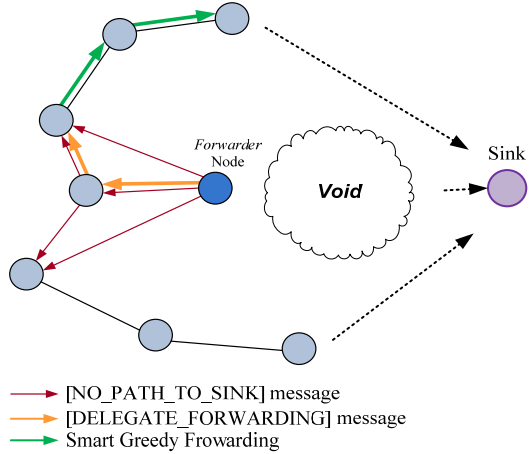

→ [NO_PATH_TO_SINK] message
→ [DELEGATE_FORWARDING] message
→ Smart Greedy Frowarding

**Figure 7: A blocking situation where a forwarder node has no neighbor closer to the sink node than itself.**

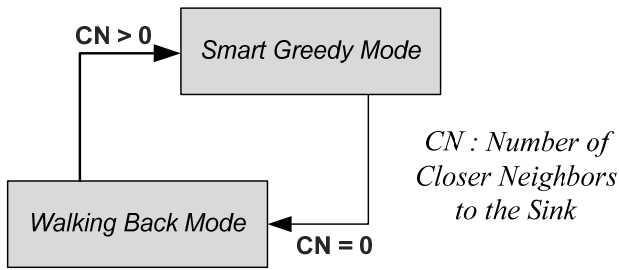

*CN : Number of Closer Neighbors to the Sink*

**Figure 8: Overview of GEAMS routing mode switching.**

IV. SIMULATION AND EVALUATION

In this paper, we have considered a homogenous WSN in which nodes are randomly deployed through the sensing field. The sensing field is a rectangular area of 500m x 200m. The sink node is situated at a fixed point in the righter edge of the sensing field at coordinates (490, 90) while a source node is placed in the other edge at coordinates (10, 90).

We consider a WSN for video surveillance. In response to an event, the source node will send images with a rate of 1 image per second during 30 seconds. The image stream is sent to the sink node for further processing or to be forwarded to a control center situated in another network making the sink node acting as a gateway.

To demonstrate and evaluate the performance of GEAMS, we used OMNeT++ 4.0 which is a discrete event network simulator [10]. To prove the effectiveness of GEAMS, we have also implemented the GPSR algorithm and compared the simulation results. Table 1 summarizes the simulation environment.

We have considered that the link data is of type IEEE 802.15.4 and in which the data rate can be proportional to the transmission distance.

We have varied the network topology by varying the number of sensor nodes to obtain network of 30, 50, 80 and 100 nodes (see Figure 9, Figure 10, Figure 11 and Figure 12). We consider the minimum distance between two neighbors node greater than 1 meter.

| Parameter | Value |
|---|---|
| Network Size | 500m x 200m |
| Number of Sink Nodes | 1 |
| Number of Source Nodes | 1 |
| Number of Sensor Nodes | 30, 50, 80, 100 |
| Number of Images | 30 images |
| Image Size | 10Kb |
| Image Rate | 1 image/sec |
| Maximum Radio Range | 80 meters |
| Link Data Rate | $250\ Kbps/\sqrt{Link\_Length}$ |

**Table 1: Simulation parameters.**

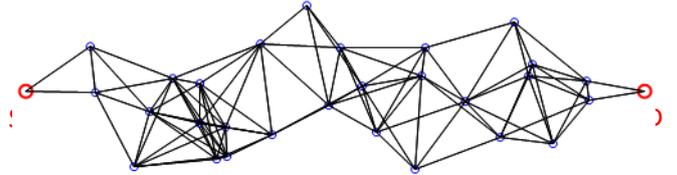

**Figure 9: A 30 nodes network topology.**

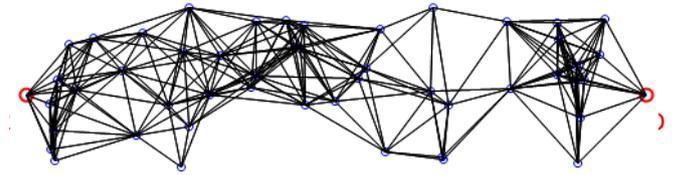

**Figure 10: A 50 nodes network topology.**

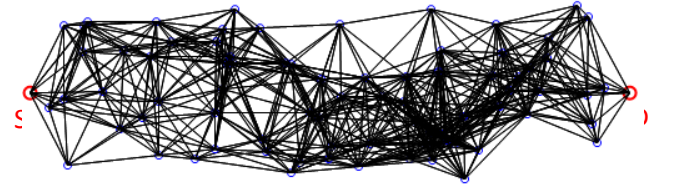

**Figure 11: A 80 nodes network topology.**

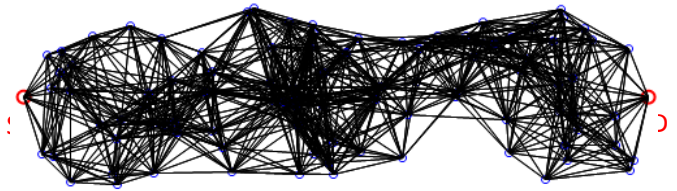

**Figure 12: A 100 nodes network topology.**

For each topology, we have measured various parameters: the number of dead nodes (see Figure 13), the distribution of the network energy using the mean value (see Figure 14) and the variance (see Figure 15), the distribution of mean energy consumption by partitioning the network into regions of 40 meters width (see Figure 16, Figure 17, Figure 18 and Figure 19), the distribution of the packet transmission delay using mean value (see Figure 20) and variance (see Figure 21), and finally the number of lost packets (see Figure 22).

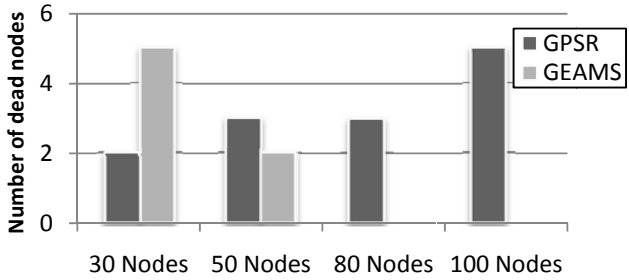

**Figure 13: Number of dead nodes in the different topologies.**

*We can clearly see that as the number of nodes within the network increases, the number of dead nodes in the case of GEAMS protocol tends to zero. However, GEAMS seems to be not adapted in less dense network.*

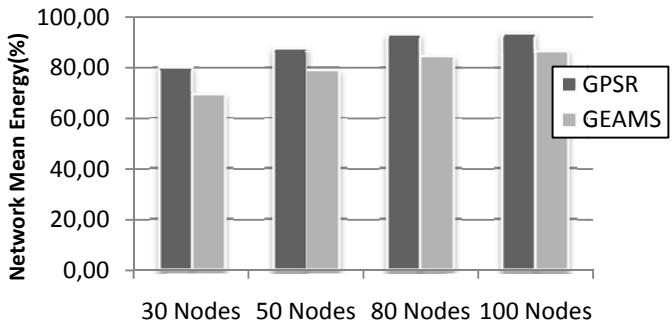

**Figure 14: Mean energy of the network in the different topologies.**

*In the case of GEAMS and by using more nodes to route packets towards the sink, the total remaining energy in the network is lesser than in the case of GPSR.*

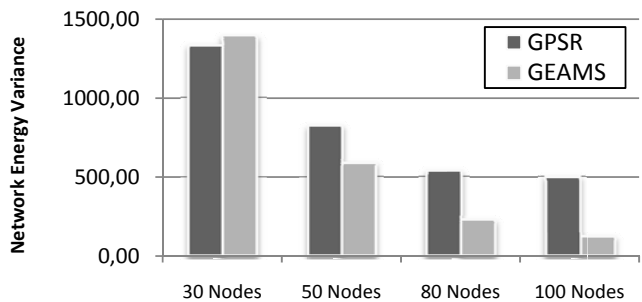

**Figure 15: Network energy variance.**

*By using few nodes to transmit packets, GPSR leaves more nodes unused and so leads to non-uniform energy consumption within the network. This is reflected by the bigger variance compared to the energy consumption in the case of GEAMS.*

*Global energy distribution:*
Because of the inflexible selection of the next forwarder node, the GPSR keeps various nodes unused and utilizes a few nodes for sending packets. This explains that GPSR mean energy is being bigger than in the case of GEAMS protocol. However, the energy distribution in the network is well distributed with GEAMS compared to GPSR since most of the nodes can be active due to multipath routing.

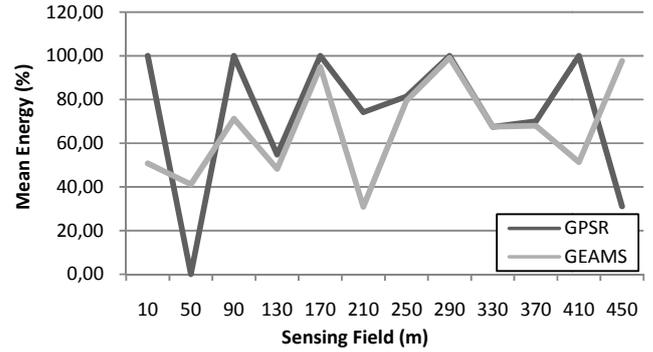

**Figure 16: Mean energy by regions of 40meters in a 30 nodes network.**

*GPSR routing decision can lead to network partitioning into completely disconnected sub networks. This case is shown in region [50, 90] which shows node dying.*

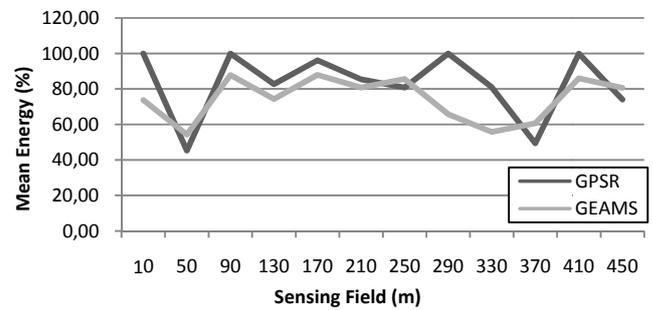

**Figure 17: Mean energy by regions of 40meters in a 50 nodes network.**

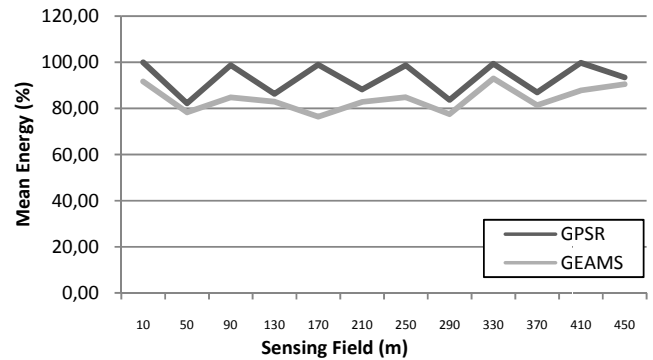

**Figure 18: Mean energy by regions of 40meters in a 80 nodes network.**

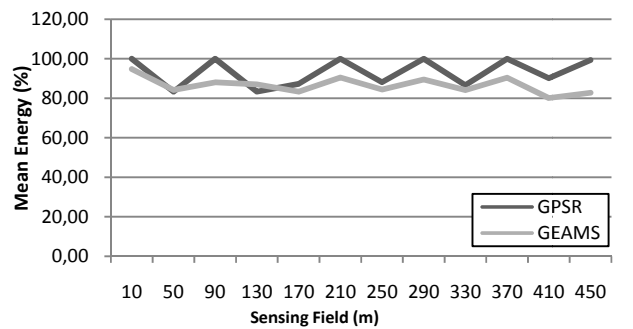

**Figure 19: Mean energy by regions of 40meters in a 100 nodes network.**

*Local energy distribution:*
Figure 16, Figure 17, Figure 18, and Figure 19 illustrate the mean energy of the network partitioned in region of 40 meters width. We can clearly see that the energy is uniformly consumed through the network when using GEAMS routing protocol compared to GPSR routing protocol. The benefit of such a feature is preventing the network from being partitioned into sub networks completely disconnected if some nodes died before the other (case of GPSR).

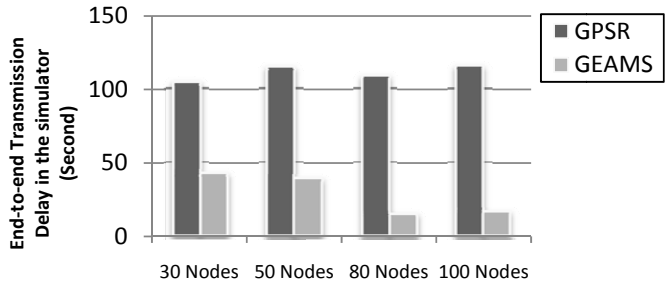

Figure 20: Mean packet transmission delay.

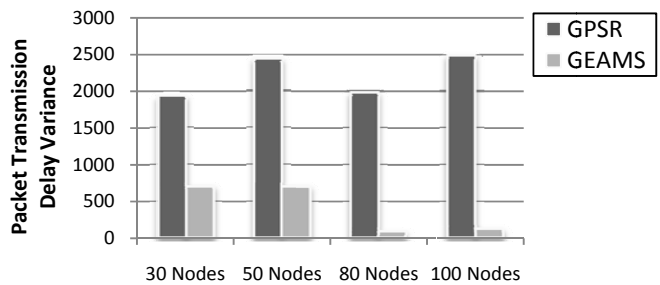

Figure 21: Packet transmission delay variance.

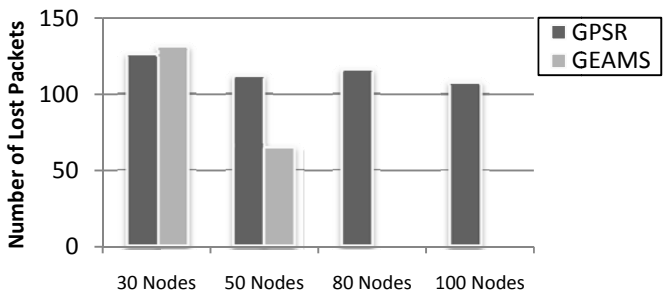

Figure 22: Number of lost packets.

*Packet loss and Transmission delay:*
Because of the use of multiple paths to transmit data packets, the packet transmission delay has been extremely decreased. The packet loss has also been decreased. This enhancement can be explained by the following points:
− The use of the same path will increase the time spent inside the buffers (queue) among this path which leads to a traffic congestion.
− Packet loss may occur because sensors cannot keep packets for a long time in its buffers and this is due to the hard resource constraint.

These results demonstrate clearly the ability of GEAMS to deliver multimedia traffic (Images traffic in our case) and enhancing the QoS compared to GPSR (lowering the end-to-end delay and packet loss ratio). GEAMS is also more suitable to dense network in which different paths to destination may exist.

V. CONCLUSION

In this paper, we have described a new algorithm namely GEAMS that is suitable for transmitting multimedia streaming over WMSNs. Because nodes are often densely deployed, different paths from source nodes to the base station may exist. To meet the multimedia transmission constraints and to maximize the network lifetime, GEAMS exploits the multipath capabilities of the WSN to make load balancing among nodes. Simulation results compared to GPSR show that GEAMS is well suited for WMSNs since it ensures uniform energy consumption and meets the delay and packet loss constraint.